\documentclass[prb,aps,preprint,showpacs, lengthcheck,showkeys, longbibliography]{revtex4-1}
\pdfoutput=1
\usepackage{amsmath}
\usepackage{amssymb}
\usepackage{graphicx}

\begin{document}

\title{Examining the Location of the Magnetopause in an Undergraduate Lab}

\author{James P. Crumley}
\author{Ari D. Palczewski}
\author{Stephen A. Kaster}
\affiliation{Department of Physics,
College of Saint Benedict / Saint John's University,
Collegeville, MN 56321}

\email{jcrumley@csbsju.edu}
  \homepage{http://www.physics.csbsju.edu/~jcrumley/}
\date{\today} 

\begin{abstract}
Integrating areas of current research into undergraduate physics labs can
be a difficult task.  The location of the magnetopause is one problem that
can be examined with no prior exposure to space physics.  The magnetopause
location can be viewed as a pressure balance between the dynamic pressure
of the solar wind and the magnetic pressure of the magnetosphere.  In this
lab sophomore and junior students examine the magnetopause location 
using simulation results
from BATS-R-US global MHD code run at NASA's Community Coordinated Modeling
Center.  Students also analyze data from several spacecraft to find
magnetopause crossings.  The students get reasonable agreement
between their results and
model predictions from this lab
as well as exposure to the tools and techniques of space physics.
\end{abstract}


\maketitle

\section{\label{sec:intro} Introduction}

Physics laboratory courses often emphasize experiments that relate directly
to physical concepts being covered in lecture courses that students
are taking concurrently.  These experiments are intended to allow students
to see for themselves how physical concepts work in the real world.
Experiments of this sort tend to deal with physics that has been understood
for some time and often involve performing classic experiments of the past.
Experiments of this type do add to student understanding, but a steady diet
of them can leave students disconnected from the 
current practice of physics. Maintaining the students' intellectual
curiosity about science is a key to retaining students in science majors.
\cite{Seymour:2000}  One way to maintain that curiosity is to create
experiments throughout the curriculum that
expose students to areas of current research to supplement
classical experiments. 

Designing experiments that deal with areas of current research and
are accessible to undergraduate students can be a difficult task.
This task is particularly difficult for space physics because most students 
have had little exposure to space and plasma physics.\cite{Stern:1997},
Space physics
applications are often complex, defying simple treatment, and they often
rely on advanced electricity and magnetism, which students often have late
in their undergraduate curriculum.

In this paper we will discuss our use of the location of Earth's 
magnetopause as a topic for a lab for sophomore or junior physics majors.
The magnetopause is defined as the boundary between Earth's magnetic field
and the interplanetary magnetic fields.  Although some topics in space
physics are quite complicated, others can be discussed at a level
appropriate for junior physics majors, or even an introductory
college course.
The magnetopause is an appealing topic for introducing
space physics,  since the magnetopause can be introduced 
at a fairly elementary level as a pressure balance,  after which
more complicated models can also be examined.  This lab involves
examining the magnetopause location using computer simulations and spacecraft
data.

In Section~\ref{sec:magnetosphere} we introduce some background information
regarding the magnetosphere in general, and the magnetopause in particular.
The portion of the lab using computer simulation to examine the
magnetopause location is discussed in Section~\ref{sec:simulation}, while 
the portion of the lab using spacecraft observations is discussed in 
Section~\ref{sec:data}.  In Section~\ref{sec:disc} the process of developing this 
magnetopause lab and its evolution over eleven
years of use are examined. Conclusions as well as ideas about similar sorts
of labs are examined in Section~\ref{sec:conclusion}.

\section{\label{sec:magnetosphere} Magnetosphere Introduction}

\subsection{The Magnetosphere and the Magnetopause}

The structure and behavior of the area of space just outside of the Earth's 
neutral atmosphere is a problem that physicists have worked on for quite
some time. The first clues to the nature of this region
came from observations of the magnetic field on
the Earth and observations of comets.\cite{Russell:1995a}
Though the Earth's magnetic field had been used since ancient times for
navigation, in 1600 William Gilbert was the first to propose that the 
Earth was a giant magnet. \cite{gilbert:1600}
Serious observations of the Earth's magnetic
field in the 1700s led to the discovery of
variations in the magnetic field of the Earth called magnetic storms.
\cite{Kallenrode:2001}  Magnetic storms were long hypothesized to be
due to the Sun,\cite{Carrington:1860}  but it took some time to arrive at a
suitable mechanism \cite{Parker:1958} for the Sun to be affecting the
magnetic field of the Earth.  A breakthrough was made when scientists realized 
that comets had two tails: one
caused by light from the Sun, the other by a stream of particles from the
Sun.\cite{Biermann:1951, Hoffmeister:1943}
This stream of particles is called the solar wind. Particles from the solar
wind, along with some escape the atmosphere, fill the nearby region of 
space with plasma. This region is now known as the
magnetosphere because the Earth's magnetic field dominates behavior
of that plasma. Magnetic storms result from complicated interactions between the
Earth's magnetosphere and the solar wind.

\begin{figure}[ht]
  \begin{center}
  \includegraphics[width=.45\textwidth]{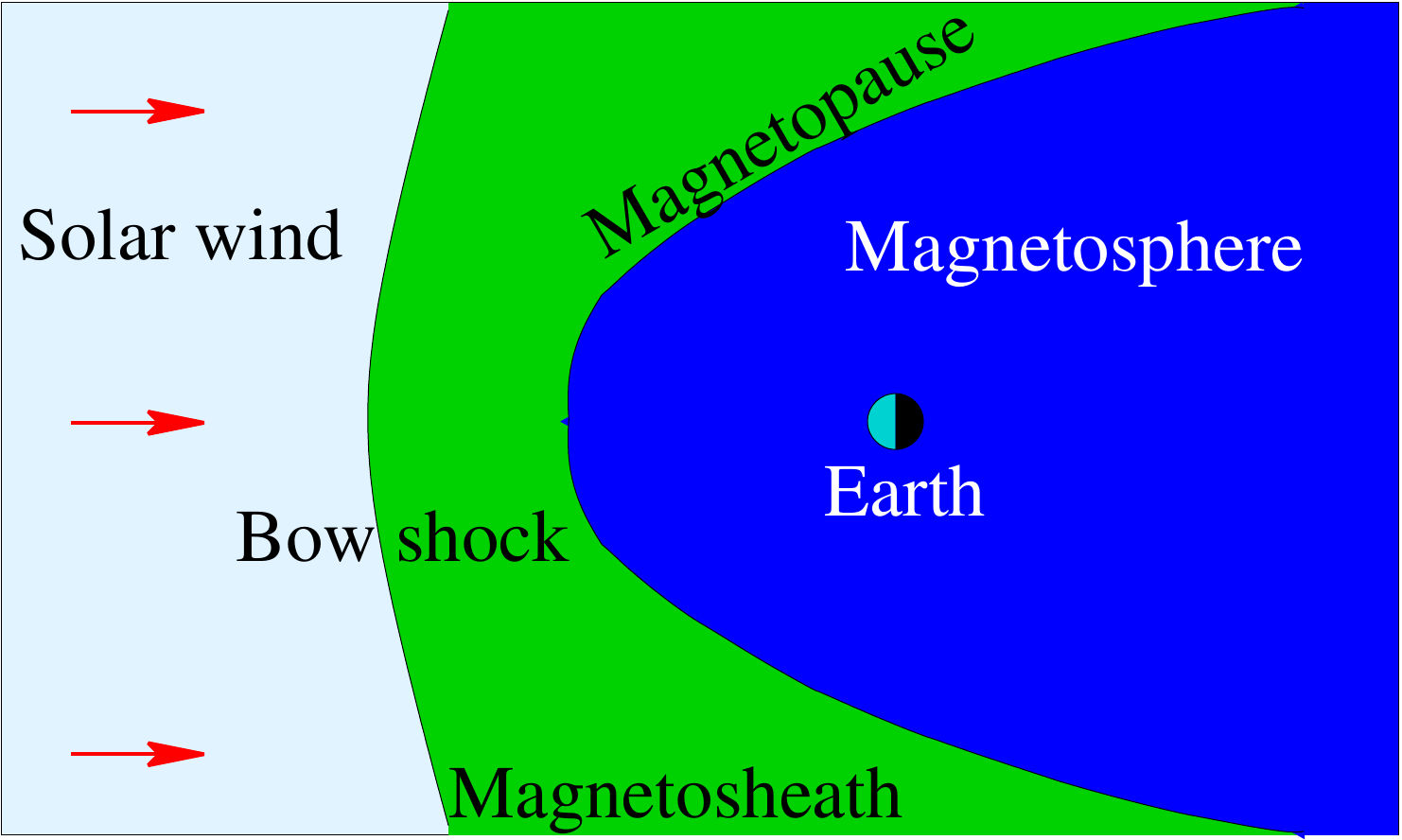}
  \end{center}
  \caption{\label{fig:magnetosphere}
   The Earth's magnetosphere.
   }
\end{figure}

The magnetic field of the Earth acts as an obstruction to the solar wind,
causing a shock called the bow shock and a boundary region called the
magnetosheath  (see Figure~\ref{fig:magnetosphere}). The bow shock forms
because the solar wind is supersonic and it hits a n obstruction in the
form of the Earth.  The resulting bow shock is similar to the shock caused
by a jet travelling faster than the speed of sound.
Besides causing the bow shock, the interaction with
the solar wind compresses the Earth's magnetic field on the day side, 
and stretches it on the night side, 
resulting in the asymmetrical shape of the magnetosphere.  As a first
approximation, the location of the magnetopause is set by the balance
between the dynamic pressure of the solar wind and the magnetic pressure of
the magnetosphere.  Variations in the solar wind lead to movement of the
the magnetopause location, as well as variations in the magnetic field
measured on Earth.  Currents within the magnetosphere and ionosphere also
affect the magnetic field measured on Earth. \cite{Parks:2004}

The existence of a magnetosphere with a magnetopause of this sort was
first posited to help explain variations in the readings of Earth-bound 
magnetometers \cite{Chapman:1931} and to tie those variations to processes
on the Sun. After these predictions, early spacecraft missions confirmed
the existence of both the magnetosphere and the magnetopause.
\cite{Cahill:1963}  Since  that time, there have been advancements in the
observation, theory, and modeling of the magnetopause location.

\subsection{Magnetopause Pressure Balance}

In simple models, the magnetic field of Earth is treated as a dipole:
\begin{equation}
  B_{dipole} = B_o\biggl(\frac{\text{R}_\text{E}}{r}\biggr)^3,
  \label{eqn:dipole}
\end{equation}
where $B_o$ is the surface magnetic field at the equator, R$_\text{E}$ is
the radius of the Earth, and $r$ is the distance from the
center of the Earth to the location of interest. In the ionosphere
and magnetosphere,  the magnetic field is caused by plasma currents, as
well as Earth's magnetic field.  

\begin{figure}[ht]
  \begin{center}
    \includegraphics[width=.45\textwidth]{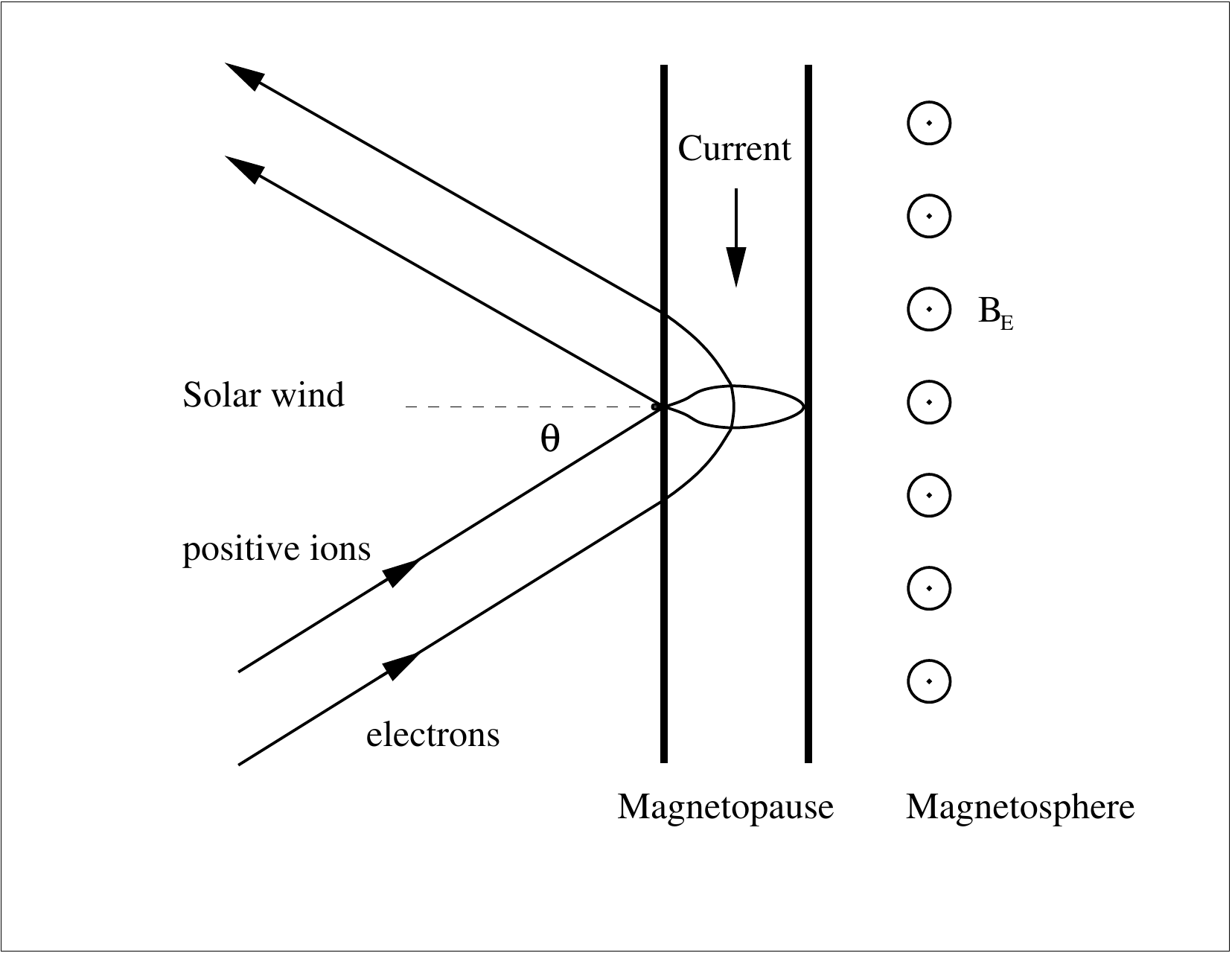}
  \end{center}
  \caption{\label{fig:pressure_balance}
   The pressure balance at the magnetopause between the solar wind 
   dynamic pressure and Earth's magnetic pressure. The electrons
   rotate counter-clockwise and the positive ions rotate clockwise
   when reflecting at the magnetopause, driving the Chapman-Ferraro 
   current. Based on similar diagrams from Willis\cite{Willis:1971}. }
\end{figure}

As mentioned above, the location of the magnetopause can be thought of as a
 pressure balance between the dynamic pressure of the
solar wind and the magnetic pressure of Earth (see
Figure~\ref{fig:pressure_balance}):
\begin{equation}
   2 \rho_{sw} (v_{sw} \cos{\theta})^2 = \frac{1}{2\mu_0} B_{inside}^2.
   \label{eqn:pressure_balance}
\end{equation}
where  $\rho_{sw}$ is the mass density of the solar wind,   $B_{inside}$
is the magnetic field inside the magnetopause, and $\theta$
is the angle of incidence of the solar wind (the
$v_{sw} \cos{\theta}$ term is the solar wind speed
normal to the magnetopause).

Due to the compression of the magnetosphere
$B_{inside}$ is not simply the dipole magnetic field of the Earth.
In the Chapman-Ferraro Model \cite{Chapman:1931} for the magnetopause
boundary location, there is a current that runs tangential to the
magnetopause boundary in the dawn to dusk direction. The Chapman-Ferraro
current results from the Lorentz force causing
electrons and positive ions to rotate opposite
directions when they reflect at the magnetopause (see
Figure~\ref{fig:pressure_balance}).  This current
causes a magnetic field that cancels the magnetic field of the Earth just
outside the magnetopause, and doubles the magnetic field inside the
magnetopause \cite{Parks:2004} so that:
\begin{equation}
  B_{inside} = 2 B_{dipole} = 2  B_o\biggl(\frac{\text{R}_\text{E}}
  {r}\biggr)^3.
  \label{eqn:Bdouble}
\end{equation}

Combining Eqs.~(\ref{eqn:pressure_balance}), (\ref{eqn:dipole}), and
(\ref{eqn:Bdouble}) and solving for the location of the magnetopause
leads to:
\begin{equation}
  \frac{r}{\text{R}_\text{E}} = \biggl(\frac{B_o^2}{\mu_o \rho_{sw} (v_{sw}
    \cos^2{\theta})}\biggr)^{\frac{1}{6}}.
  \label{eqn:magploc}
\end{equation}
Assuming that solar wind consists of protons and electrons coming in 
normal to the magnetosphere and 
substituting in typical values of 10 protons/cm$^3$ and 400 km/s, 
Eq.~(\ref{eqn:magploc}) gives a distance to the magnetopause subsolar point, 
r$_o$, of roughly 8~ R$_\text{E}$.  (The subsolar point is 
the location on the magnetopause 
along the line from the Earth to the Sun.)  The observed value
is for those conditions is  10~R$_\text{E}$.  

While the Chapman-Ferrari model is a good start, a
more thorough empirical expression derived from research which includes
several factors ignored above and takes $\theta =0$ gives
\begin{equation}
  r_o(\text{R}_\text{E}) = 107.4 \bigl(n_{sw} v^2_{sw}\bigr)^{-\frac{1}{6}}.
  \label{eqn:standoff}
\end{equation}
Eq.~(\ref{eqn:standoff}) accounts for factors including
the presences of positive ions heavier than protons in the solar wind,
and the magnetic field strength at the subsolar point
is not $2 B_{dipole}$, but is instead $2.44 B_{dipole}$.
In Eq.~(\ref{eqn:standoff}), $r_o$ is the distance from the center of 
Earth to the magnetopause subsolar point in R$_E$, 
$n_{sw}$ is the number density of the plasma in the
solar wind in cm$^{-3}$, and $v_{sw}$ is the speed of the solar wind in
km/s.\cite{Walker:1995}  Note that this expression has the same dependence on
solar wind speed and number density as Eq.~(\ref{eqn:magploc}), but a 
different leading constant, so Eq.~(\ref{eqn:standoff}) is still closely
related to the Chapman-Ferra model.

%

\subsection{Recent Magnetopause Modeling}

Much work has been done on modeling the location and shape of the
magnetopause since Chapman and Ferraro's original model. \cite{Chapman:1931}
Along with the dynamic pressure of the solar wind, it was found that
the orientation of the interplanetary magnetic field (the magnetic
embedded in the solar wind) plays a 
key role in determining the shape of the magnetopause.
\cite{Fairfield:1971}  More recent studies have concentrated on using
databases of magnetopause crossings of various spacecraft and the
solar wind conditions at those times to formulate
empirical expressions for the magnetopause location. \cite{Sibeck:1991,
Petrinec:1996}  In the lab described here
we use fits from Ref.~\onlinecite{Shue:1998}
in the spacecraft data portion of the lab to predict the magnetopause
location.
\begin{eqnarray}
  \label{eqn:shue}
   r= r_o \Bigl(\frac{2}{1 + \cos{\theta}}\Bigr) ^\alpha  \\
   r_o = (10.22 + 1.29 \tanh{[0.184(B_z + 8.14)]})(D_p)^\frac{-1}{6.6} \\
   \label{eqn:shue3}
   \alpha = (0.58 - 0.007 B_z) [ 1 + 0.024 \ln{(D_p)}] 
\end{eqnarray}
In Eqs.~(\ref{eqn:shue})--(\ref{eqn:shue3}), $r$ is the distance from Earth to the magnetopause boundary
in R$_\text{E}$,
$r_o$ is the distance from Earth to the subsolar point of the
magnetopause in R$_\text{E}$,  $B_z$ is the
$z$-component of the solar wind's magnetic field in nT, 
$D_p$ is the dynamic pressure of the solar wind in nPa, and $\alpha$ is a
unitless number representing the
amount of tail flaring on the night side of the magnetosphere.  The tail 
flaring describes the shape of the magnetopause (See
Figure~\ref{fig:magnetosphere}).  Small values of $\alpha$ lead to closed,
ellipsoid-like magnetospheres topologies, 
while larger values large values lead to open magnetospheres.

\section{\label{sec:simulation}Magnetopause Location from Simulation}

In this portion of the lab  students perform computer simulations using various
solar wind conditions and determine where the subsolar point of the
magnetopause is from their results.  These results are fit to an expression
of the form of Eq.~(\ref{eqn:standoff}), with the leading constant 
as a free parameter.  The students compare their constant to the value of 
107.4 from Eq.~(\ref{eqn:standoff}).

\subsection{Simulation Environment}

The students model the magnetosphere by running the BATS-R-US
(Block-Adaptive-Tree-Solarwind-Roe-Upwind-Scheme)
\cite{Powell:1999, Gombosi:2001, Toth:2005} simulation on
supercomputers at the Community Coordinate Modeling
Center (
{CCMC}).\cite{CCMC:2015}
BATS-R-US solves the 3D magnetohydrodynamics
equations in finite volume form using Roe's Approximate Riemann Solver on an
adaptive grid. The simulation is run on NASA supercomputers at the CCMC 
where simulation runs can be requested by the public.  Results from the
simulations can be explored and visualized through the use of a standard
web browser. The documentation provided  is quite thorough, so explaining
to students how to request a run and access the results is relatively
easy.  The resources that
CCMC provides allow for access to 
powerful simulations without the need to have local access to 
supercomputer hardware or to install and maintain simulation codes.

\subsection{Finding the Magnetopause Location in Simulation Results}

In the process of finding the subsolar point of the magnetopause in their 
simulation results, each lab group was required to come up with their
own standards for determining the magnetopause location.  
The lab materials explain what processes will be going on near the magnetopause
and how that might affect the students' graphs, but no prescription for
finding the magnetopause is given. The students are also encouraged
to ask their instructor for assistance if they have difficulties
defining where the magnetopause is in their results.
The task of defining their own standards serves
several purposes.  The students are stimulated by this requirement to
examine their graphs more closely than they would be if a prescriptive method 
for finding the magnetopause were given. Furthermore, the openness of this 
task is a fair reflection of real lab work 
since interpretation of simulation results often requires  discretion.

The instructions regarding finding the magnetopause location advise
the student to look at more than one plasma parameter when finding the
magnetopause, since the magnetopause should be evident in more than
one type of data.  Since the magnetopause is defined as the boundary 
between where the Sun and the Earth's magnetic fields dominate, it is
clear that a signal should be visible in the magnetic field
results.  The magnetopause
also affects the motion of the plasma particles, so differences
in the plasma bulk velocity and number density should also be visible.

\begin{figure}[ht]
  \begin{center}
  \includegraphics[width=.5 \textwidth]{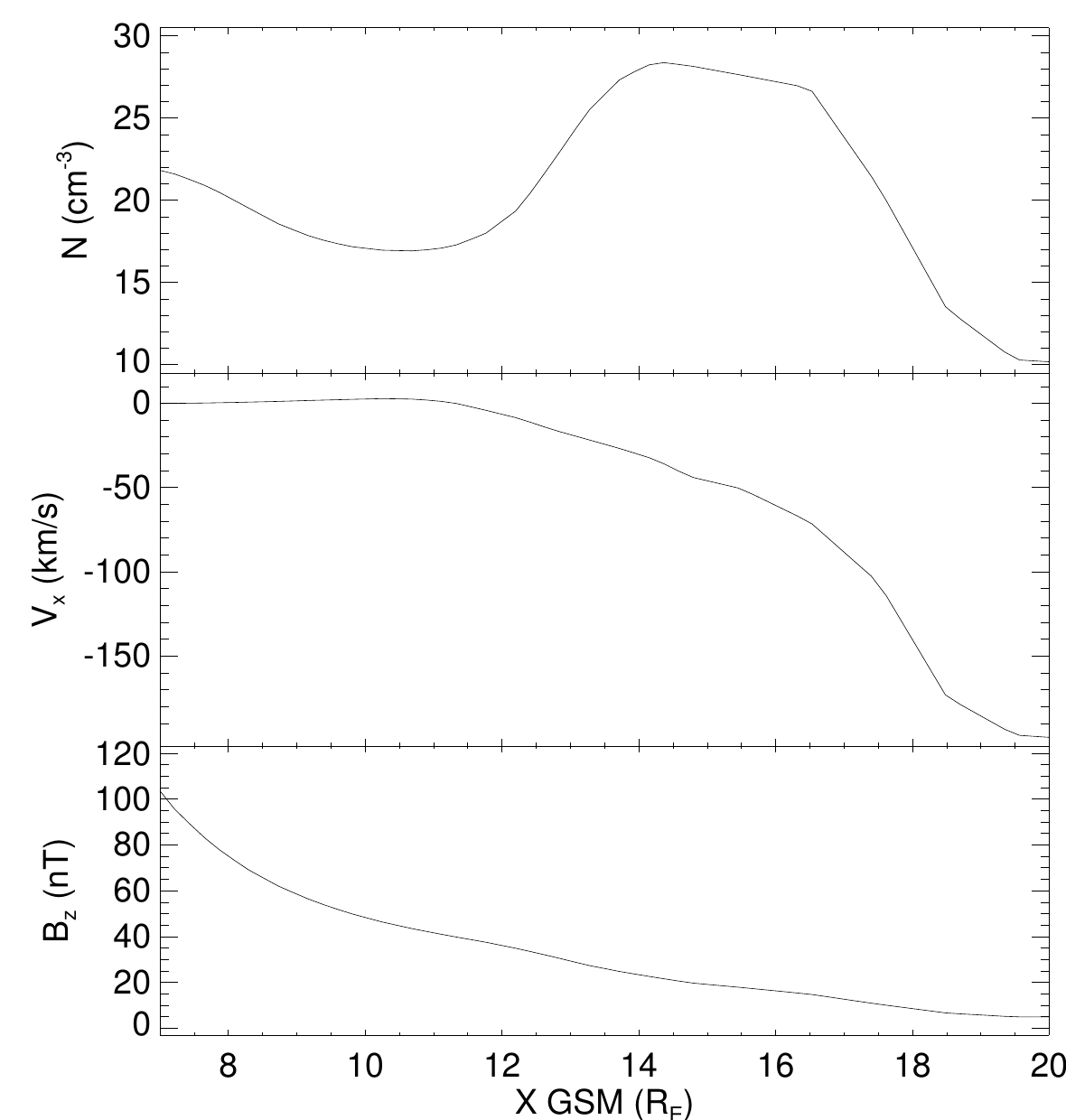}
  \end{center}
  \caption{\label{fig:batrus}
    Simulation results plot used to find subsolar point of the
    magnetopause.  The data are from 
     five minutes into a typical
    simulation run and they are plotted along the line from
    the Earth to the Sun.  The top panel shows the number
    density, the middle panel shows the x-component of the plasma flow
    velocity, and the bottom panel shows 
    z-component of the magnetic field.
    Geocentric solar magnetospheric (GSM) coordinates
    are used here. In this system the x-axis points from the Earth to
    the Sun, the z-axis points in the direction of Earth's north
    magnetic pole, and the y-axis completes the right-handed
    system. \cite{Kivelson:1995}}
\end{figure}

Figure~\ref{fig:batrus} shows an example of simulation data used by students to find the
subsolar point of the magnetopause. On this plot, the magnetopause is
located at roughly 11 R$_E$. In the number density plot, the bump in the number 
density corresponds to the buildup of plasma in the magnetosheath, so
the inner boundary of that bump corresponds to the magnetopause.  In the 
plot of the x-component of the velocity, the magnetopause is seen as the
location where the value goes to 0, since the plasma from the solar wind is
diverted around the magnetosphere at the magnetopause.  Finally, in the plot
of the z-component of the magnetic field, there is an almost imperceptible
shift at 11 R$_E$. So in this case,  it is easier to
find the magnetopause in the plasma results for the simulation than in
the magnetic field results.

Since the development of this lab, BATS-R-US has added the magnetopause
location as a parameter that the simulation calculates itself.  The
availability of the simulation calculation of the magnetopause has been a good
check on the students' estimates of the magnetopause location.  The students
compare their estimates to the simulation values for the magnetopause
location and comment on whether the differences between the two sets of
data are systematic.

\subsection{Empirical Fit}

\begin{figure}[ht]
 \includegraphics[width=.5\textwidth]{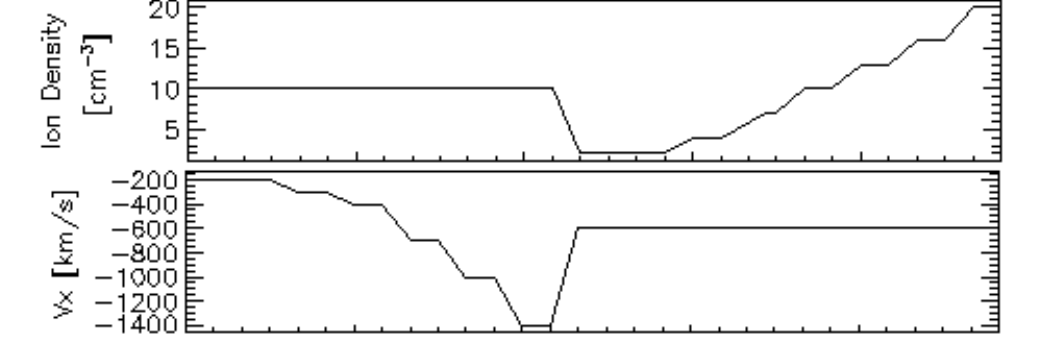}
 \caption{\label{fig:sim_input}
Sample plots of the input solar wind conditions that are varied
in the simulation portion of this lab. Note that in this case
the solar wind number density (upper plot) is held constant
for the first half and varied for the  second half of the simulation, 
while the solar wind speed (lower plot) is held constant
for the first half and varied for the  second half of the simulation.
Also note that the step pattern is used in the solar wind conditions
to allow the magnetopause location to come to equilibrium after the 
conditions are varied.
 }
\end{figure}

In this lab, students find the subsolar point of the magnetopause
for a variety of solar wind conditions and fit their data to
Eq.~(\ref{eqn:standoff}), finding their own value to compare to the leading 
constant of 107.4.  The solar wind parameters for their
simulations are set to test
Eq.~(\ref{eqn:standoff}) in two different ways.
In one set, the solar wind speed is held
constant and the solar wind number density is varied, and in the other
set the fixed and varying parameters are switched (see
Figure~\ref{fig:sim_input}). 
This analysis is completed with the students' estimates of the magnetopause
and separately with the simulation's calculation of the magnetopause
location, which leads to four estimates of the leading constant
(two estimates for each half of the data).
Most groups get
good agreement between their leading constants for the fixed solar wind
speed versus fixed number density data, showing that the students
are reasonably consistent in their estimates of magnetopause location.
On the other hand, their results often do not agree to within
uncertainties for the simulation versus student estimates of the
magnetopause location. These discrepancies are typically caused by 
systematic issues in how the students are defining the magnetopause
location.
More prescriptive directions on how to find the
magnetopause location might yield better constants, but we are wary of
losing the experience that the students gain by defining their own
standards.

\section{\label{sec:data}Spacecraft Data}
In this section of the lab, students search for magnetopause
crossings for three sets of data chosen from several events
and several
spacecraft (Geotail, Polar, the GOES satellites, and the LANL 
geosynchronous satellites). The data is from well-known magnetic
storms (such as the Halloween 2013 storm \cite{Lopez:2004} shown
below), 
and the data used is publicly available online.\cite{CDA:2015}
The students first compare the actual position of the spacecraft
to the location of the magnetopause given by 
Eqs.~(\ref{eqn:shue})--(\ref{eqn:shue3}) 
to predict where the magnetopause crossings should be for 
appropriate solar wind conditions.
The  solar wind data comes from 
the ACE spacecraft \cite{Chiu:1998}, 
which orbits outside Earth's magnetosphere at the L1 point 
on the line between Earth and Sun, allowing ACE to take constant
measurements of the
solar wind upwind from Earth.  The calculations of the predicted
magnetopause
location also account for the propagation time for the solar wind to get from
the L1 point (roughly 240~R$_\text{E}$ from Earth) 
to the magnetopause (roughly 10~R$_\text{E}$ from Earth).

Next, the students examine the particle and magnetic field data from that spacecraft
for signs that the spacecraft made magnetopause crossings at the
predicted time or at other times. Finally, students compare the 
signs of magnetopause crossings that they see in the magnetic field and
particle data and discuss their perceptions of the difficulty
of interpreting those two types of data.

\subsection{Magnetopause Crossings in Spacecraft 
  Data \label{subsec:data_example}}

\begin{figure}[ht]
 \includegraphics[width=.5\textwidth]{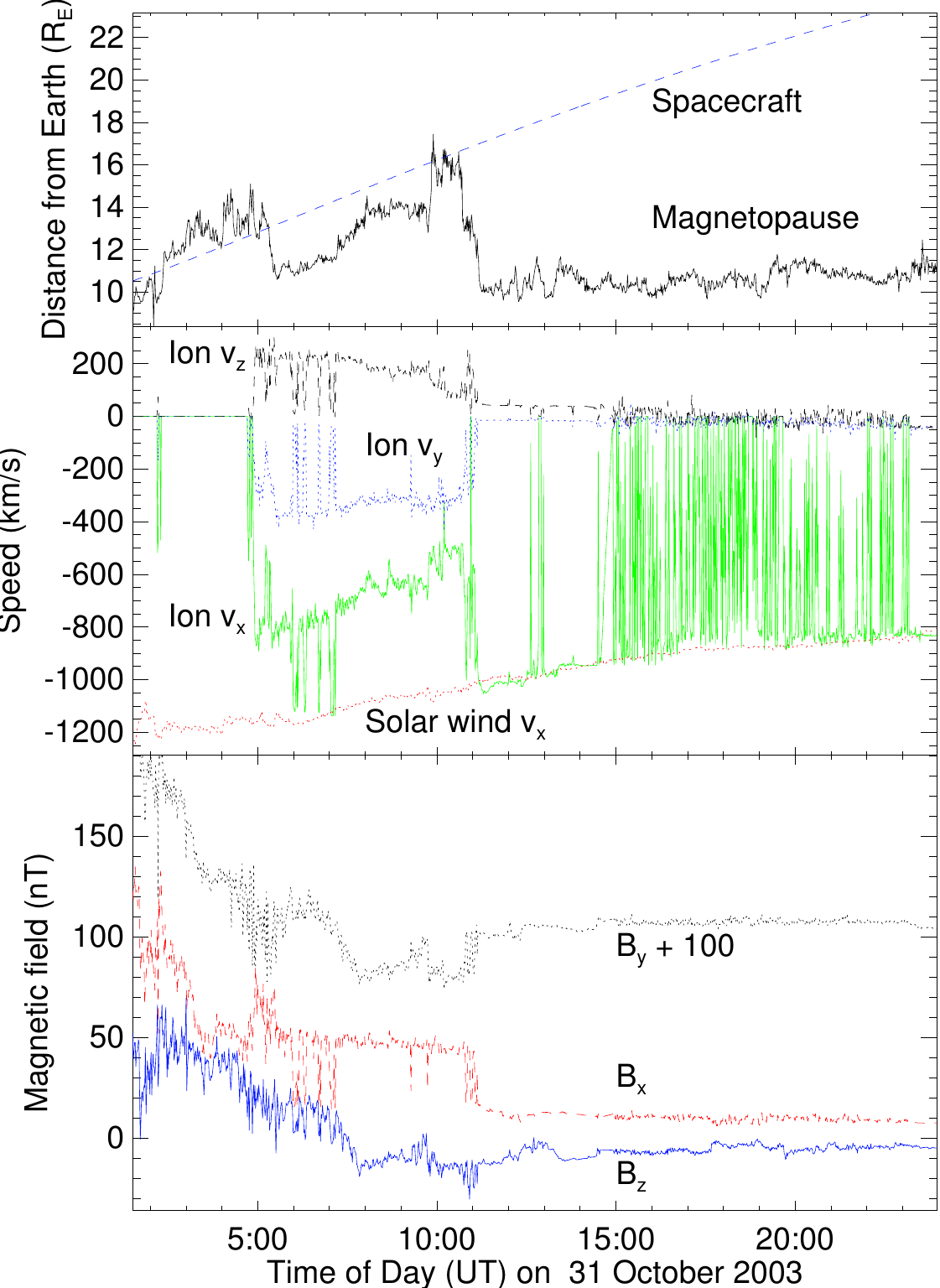} \\
\caption{\label{fig:data} 
Plots showing
Geotail's magnetopause crossings on 31 October 2003 (color online).
The top panel shows
the Geotail spacecraft's distance from Earth (solid line) 
and the predicted location of the magnetopause along the line from the Earth to Geotail (dashed line) calculated using Eqs.~(\ref{eqn:shue})--(\ref{eqn:shue3}) based on
the ACE spacecraft measurements of solar wind conditions.
The middle panel 
shows Geotail's measurements of the ion flow velocity and
the x-component of the solar wind speed measured by ACE (lowest dotted
line), all in geocentric
solar ecliptic (GSE) coordinates. In this system the x-axis points from the
Earth to the Sun, the z-axis points perpendicular to the plane
of earth's orbit, and the y-axis completes the right-handed system.
\cite{Kivelson:1995}
The bottom panel shows the three components of the magnetic field measured
by Geotail in GSE coordinates. Note that 100 nT has been added to 
B$_y$ in order separate the lines on the plot.}

\end{figure}

Figure~\ref{fig:data} shows an example of the search for spacecraft
crossings of the magnetopause, in this case Geotail, that students do in
this lab.  The data shown are from 31 October 2003.
On that day a large CME hit Earth, causing auroras that were
visible throughout much of the United States. \cite{Oler:2004, Baker:2004}

The top panel of Figure \ref{fig:data}
predicts several magnetopause crossings, but the
most notable crossings are at roughly 5:00, 10:00, and 11:00.  From 1:30--5:00
and from roughly 10:00--11:00, Geotail is predicted to be 
inside the magnetosphere. 
For most of the
rest of that day Geotail was outside the magnetosphere.  
The data in the middle and bottom panels of Figure \ref{fig:data}
show the crossing from inside to outside the magnetosphere at 11:00
and the data also shows other magnetopause crossings, though as described
below the crossings seen in the data differ somewhat from the predictions.

The middle panel of Figure \ref{fig:data}, which shows the ion flow velocity
measured by Geotail and the x-component  of the solar wind speed measured by
ACE, has three  apparent regimes. During this day the x-component of the solar
wind velocity slowly changes from $-$1200 km/s to $-$800 km/s.
From 0:00--5:00 most of the Geotail ion flow velocity data is
missing.  From 5:00--11:00 the three components of the ion
flow velocity measured by Geotail oscillate
near values of $-$700 km/s, $-$400 km/s, and 200 km/s respectively. While from
11:00--24:00 the y- and z-components oscillate near 0 km/s, and the
x-component is based at roughly $-$1000 km/s, with spikes up to almost 0 km/s.
The behavior of the x-component during this last time period
can be interpreted as being due to the 
spacecraft being just outside the magnetopause in  the
magnetosheath.  When the x-component is near $-$1000 km/s it matches
the solar wind speed, suggesting that Geotail is outside the magnetopause
at those times.  The noisiness of these measurements of the ion speed in
the magnetosheath is likely due to reflection of the solar wind ions
off the magnetopause and the movement of the magnetopause as the
solar wind conditions vary.

The velocity measurements act much differently from
5:00--11:00, suggesting that the spacecraft is inside the magnetosphere
during those times.  Notice  that during this time
there are several short time periods where all components of the data
have spikes which match
their magnetosheath values.  These results
suggest that during this time period the
spacecraft is near the magnetopause boundary and that fluctuations in the
solar wind cause the magnetopause to oscillate back and forth
across Geotail's position. 

The bottom panel, showing Geotail's magnetic field measurements, has
similar regions of behavior.  From 0:00--11:00 all three components 
have spiky measurements which trend downward. The downward trend
is due to the decreasing magnitude of the Earth's dipole field as Geotail
gets further from Earth, while the spikes are due to disturbances in the
magnetospheric plasma during this magnetic storm.  From 11:00--24:00,
the three components are steadier and oscillate near 10 nT, 5 nT, and $-$10
nT, respectively.  During this time Geotail is in the magnetosheath 
where the solar wind magnetic field dominates.  Note that during several of the
spikes in the magnetic field which occur between 0:00 and 11:00, that the
magnetic field matches the magnetosheath values.  These results support the
notion that the magnetopause oscillated back and forth past the spacecraft,
as was mentioned with the ion velocity data.

In summary, the spacecraft data (bottom two panels of
Figure~\ref{fig:data}) show that Geotail crossed the
magnetopause.  The data suggest that Geotail was inside 
the magnetopause from 0:00-11:00 (though there were some brief crossings
during that time), and that it was outside the magnetopause from
11:00-24:00. The crossings seen in Geotail's data tell
a slightly different story than what the top panel 
of Figure~\ref{fig:data} predicted.  The prediction had a more complicated
series of crossings up until a final crossing to the outside of the
magnetopause at 11:00.  Overall though, the predicted and actual crossings
agree fairly well.

\subsection{Results}

Most students get reasonable results for this section of the lab. As
seen in the example in Section~\ref{subsec:data_example},
finding the magnetopause crossings in the data can be a
complicated process, so the students do typically need a bit of guidance
as they proceed. One of the things that troubles some of the students is 
that some of the data
sets they examine contain no magnetopause crossings. Finding no crossings
disconcerts 
students since their expectation is that all data sets will have crossings.
In research there are often data sets that do not contain the 
phenomenon being searched for, so it is good to have students do
some cases of this sort.

In general, the students do a good job of finding the crossings in the 
magnetic field data, but they have more difficulty with the particle data.
These problems are to be expected since in most of the cases the ion data
is not as clear as in Figure~\ref{fig:data}.  This problem could be alleviated
somewhat by finding events where the signs of the magnetopause crossing are
clearer in the ion data, as well as by giving the students more guidance in
the interpretation of the ion data.

\section{\label{sec:disc}Lab Development and Evolution}

This magnetopause lab grew out of a desire to
replace a plasma simulation lab 
 that students thought was too
abstract and dry. Basing a lab on the magnetopause gave us the chance to
work on many of the same data analysis skills, but with a specific and
engaging topic. 
In the eleven years that we have used this lab, it
has eveloved considerably.
In this section we will discuss the process of
developing and improving this lab 
to aid other instructors who would like to develop similar labs.

First, we should mention some more details about how this lab is used.
In our department 
we have a four semester sophomore and junior lab sequence, where the
students have a one-credit lab course that meets once a week for four 
hours. Each semester students get a different lab professor
and each section of this lab typically has eight to fourteen students. 
This lab has  
been used in both sophomore and junior lab courses, but recently 
it has been in the spring semester junior course. 
Students in our labs work in groups of two (or three if there
is an odd enrollment).
In this course
students have freedom on when they work on various experiments, but
students typically work on this magnetopause lab for three to four weeks.
While groups vary, the simulation and spacecraft data portions of the lab
typically take comparable amounts of time.

Minimal computing facilities are needed for this experiment.  For the
simulation portion, the bulk of the plotting is done on the NASA web site,
\cite{CCMC:2015}
though some simple curve-fitting (which could be done
online\cite{Kirkman:2015} or with a
spreadsheet) is also required.  For the spacecraft data portion of the 
lab, our students use IDL, but any scientific plotting program or
spreadsheet could be used.

When we first developed this lab, we also wanted to show the relevance of
space physics to our students' lives.  So for the spacecraft data portion
of the lab, we have focused on large solar storm events.  Concentrating
on large storms brings up the possibility of damage to satellites,
particularly satellites in geosynchronous orbits (where
most of the satellites used for this lab orbit).  Large solar storm events
also allows us to look at data for multiple spacecraft for a single
day and consider which of them crossed the magnetopause.  During normal
solar wind conditions, it is unlikely that more than one spacecraft would
have crossed the magnetopause in a given day.  Finally, looking at big
solar storms brings in ties to the aurora borealis which are often
visible locally during these storms.  In fact, while analyzing their
spacecraft data, some of our students remembered seeing aurora on the 
night when the data was taken.

We have revised this lab over time to try to increase what
the students learn from it.
Since the background and methods
needed for this lab differ so much from other experiments, both the lab
write-up and the lab lecture for this experiment are longer than those for
the other experiments used in the same course. One strategy that
we have settled on for this lab lecture for this experiment is
to aim for concision. Some years we have been too thorough, and the students
have been overwhelmed by the avalanche of information, and their progress
on this experiment suffered because of it. Now we try to cap the background
lecture at about an hour. We explain to the students
that this experiment is complicated, so we expect them to ask  many
questions while they are working on it. In fact, for some parts of the
experiment I worry about lab groups that do not ask me questions.

Another area where the information provided is balanced
is the amount of background versus specific procedural instructions.
The students come to the lab with little knowledge of space physics, so they
must get enough background to understand what they are doing, but not so
much that they are inundated. We have settled on a level of background
that is similar to this paper --- a brief introduction to the solar wind
and magnetosphere,
followed by more detailed information on the magnetopause and the 
pressure balance that forms it.\cite{Crumley:2015mp}
One thing that has helped is that this lab now takes place in the junior
year
during the same semester when most of the students are taking electricity
and magnetism.  Though plasmas are not typically covered in any depth
in E\&M, at least the concepts of charges and fields are fresh in the
students' minds when they come across this lab.

The interplay between simulation and observation is another important 
balance to consider when developing a lab like this one.  In space physics,
simulation and observations build upon one another, and one of the strengths 
of this lab is that it exposes students to both.  In the lab
manual,\cite{Crumley:2015mp}
as in this
paper, the simulation portion of this experiment is discussed first because
the global view that is possible with the simulation flows nicely from the
global view of the magnetosphere in the introduction to this lab.
Furthermore, the simulation gives cleaner results.  Spacecraft orbits
do not go along the line between the Earth and the Sun,
so the spacecraft data is necessarily messier than the simulation results.  
Though
simulations are presented first, the two parts of this lab can be done in
any order.  In fact, we encourage the students to work on both parts in
parallel, for reasons that are both practical---it typically takes a couple
of days for simulations results to become available---and philosophical
by switching back and forth between the simulation and observation
portions of the lab, students are more likely to see the connections
between the parts.

Another difficult part of preparing students to do this lab involves
analysis of graphs.  By the time they are juniors, our 
students are skilled at fitting curves, and interpreting $\chi^2$
values.  In this lab we ask them to do more subjective analysis when 
determining the locations of magnetopause crossings in both simulation
results and spacecraft data. Our students need help developing and
applying subjective standards for where the magnetopause crossings are
happening, but again there is a balance.  We do not give them standards,
because we want them to develop their own, but our students do need guidance
since this topic is so new to them.  

Within the simulation portion of the lab, one facet of the lab that 
has evolved
a great deal is how we recommend the students vary the solar wind speed
and number density.
 The first time students tried this lab, we gave each group
ranges for those parameters, but then let the students
decide how they wanted to vary them. Unsurprisingly, that was a little too
much freedom for the students and their results suffered.  Using
this simulation to probe the correlation between solar wind dynamic 
pressure and magnetopause standoff distance is a relatively constrained
problem, compared to looking at real changes in solar wind conditions.
The obvious option of having the solar wind parameters vary linearly 
with time leads to poor results because it takes time for changes in the 
solar wind conditions to propagate throughout the magnetosphere.  So
to improve results, students must leave the solar wind conditions the
same for multiple time steps, allowing the magnetopause location to
reach an equilibrium for a given set of conditions before the conditions
are again varied. Using this stair-step pattern, as shown in 
Figure~\ref{fig:sim_input}, for input conditions
has led to much better results, though over time
the instructions on how to create the stair-step pattern have evolved.

\section{\label{sec:conclusion}Conclusion}

We have presented a lab in which students examine
the Earth's magnetopause using simulations and spacecraft
data.  In this lab students are challenged by exposure to an active area of
research that they might not otherwise  encounter as
undergraduates.  Furthermore, not only are they exposed to space physics,
they work on a problem using real data and tools in a manner that is
not far removed from what current researchers do.  Experiences of this sort
are key in retaining physics majors and in helping physics students
determine what they want to do with their physics education.

The outlook for future space physics and astronomy labs 
involving nearly current research topics is bright.
The move toward open access to spacecraft data, coupled with the 
increase in the number of operational spacecraft taking science data,
continues to broaden the areas of space that anyone with an internet connection
can access.  Since we first developed this lab, citizen science efforts 
in general,\cite{Bonney:2009} as well as those specifically
using spacecraft data, have abounded\cite{Zooniverse:2015} 
and taking advantage of
these data resources can enrich the science classroom as well.
On the simulation side, the development of efforts to
give public access to supercomputers, along with the steady increase in
computing power, 
continues to expand the number of problems that 
anyone can find the resources to simulate.  The key difficulty to applying 
these new observational and computational powers to the teaching lab 
remains formulating appropriate problems for students to examine.
We hope that this paper sparks ideas for other problems to tackle.

\bibliography{big}

\bibliographystyle{ajpdjg}

\begin{acknowledgments}
We would like to thank Kristi Keller from NASA Goddard Spaceflight Center.
Conversations with Kristi led to the development of this lab.  Simulation
results have been provided by the 
Community Coordinated Modeling Center\cite{CCMC:2015}  at
Goddard Space Flight Center through their public Runs on Request system.
The CCMC is a multi-agency partnership
between NASA, AFMC, AFOSR, AFRL, AFWA, NOAA, NSF and ONR. The BATS-R-US
Model was developed by Tamas Gombosi et al. at the 

Center for Space
Environment Modeling\cite{CSEM:2015} at the University
of Michigan. Spacecraft key parameter data provided by 
{CDAWeb}.\cite{CDA:2015}  Data shown here from Geotail MGF
provided by S. Kokubun at STELAB, Nagoya University, and Geotail CPI
provided by L. Frank at University of Iowa.  Solar wind data shown here is
from ACE SWEPAM provided by D. J. McComas from SWRI and ACE MFI provided by
N. Ness  from Bartol Research Institute.  This research was supported by an
award from Research Corporation.
\end{acknowledgments}


\end{document}